\documentclass[7pt,
twocolumn,
prd,aps,amssymb,amsmath,tightenlines]{revtex4}
\usepackage{graphics}
\usepackage[pdftex]{graphicx}

\usepackage{amsfonts}
\usepackage{amssymb}

\newcommand{\ket}{\rangle}
\newcommand{\bra}{\langle}

\begin{document}
 %  \begin{multicols}{2}   
   \preprint{}

\title{The Helstrom Bound}
%\title{A Reevaluation of the Helstrom bound}
\author{Bernhard K. Meister}
\email{bernhard.k.meister@gmail.com}
%\affiliation{...., Imperial College, London SW7
%2BZ, UK}
\affiliation{ Department of Physics, Renmin University of China, Beijing, China 100872}

\date{\today }

\begin{abstract}
\vspace{-.298cm}
Quantum state discrimination between two wave functions  on a ring is considered. The optimal minimum-error probability 
%for the state discrimination
is known to be given by the Helstrom bound. A new strategy is introduced by inserting  instantaneously two impenetrable barriers dividing the ring into two  chambers. 
In the process, the candidate wave functions, as the insertion points become nodes,  get entangled with the barriers and can, if judiciously chosen, be distinguished with smaller error probability.
%An insertion  at a non-nodal point leads to  %an energy-dependent entanglement between the wave function and the barriers. %The modified wave function is now expressed as a sum of energy eigenfunctions. %created out of the intervals between the barriers.
 %The    wave function is modified by
%  a transfer of energy from the barrier to the  wave function,  independent of its initial shape, %from the barrier at the non-nodal point  
  % as the insertion point  becomes a node.  
% if the insertion point is not a node of the initial wave function. The barrier becomes entangled with the wave function in the process. %into the square well.
% modifying the wave-functions.
%The modified candidate wave functions, if judiciously chosen,  are now  in an energy-dependent entanglement with the barriers  and can be %, if the insertion points and the candidate wave functions are judiciously chosen,  
%distinguished with a smaller error probability. 
As a consequence, the Helstrom bound  under  idealised conditions can be violated.   %The energy required to insert the barrier is  dependent on the initial state. This enables the experimenter to gain additional information beyond the standard probing of the state envisaged by Helstrom and to improve the distinguishability  of the states. It is shown that under some conditions %in some cases
\end{abstract}
%\vspace{-.598cm}
%\pacs{PACS number(s): 0???3.65.BZ, 05.70.Ce, 05.30.Ch}
\maketitle

%{\it Keywords:}   Helstrom Bound; Bayes Cost;  Quantum Information theory 
%Quantum measurements
%Negative Result Measurements; 

\vspace{-.598cm}
\section{Introduction}
\label{sec:1a}
\vspace{-.438cm}

\noindent
Experimental design and data analysis are common challenges in science, and particular acute in quantum mechanics. %, due to recurring questions about foundational issues and restrictions on measurements.
%The optimal observational strategy has to be designed before the data is obtained. In the process one needs to know what can be measured and how the data obtained is analyzed.
%Even in an idealized situation of a one-dimensional well as discussed in this paper this
In the literature different approaches are discussed. 
%For example, there is the information theoretic approach, where one maximizes the mutual information, and the minimax approach, where one minimizes the maximum cost of a set of strategies assuming a perfidious opponent. Both methods are popular, but
Here the Bayes procedure for state discrimination with the aim to minimise the expected cost is employed. The existence of a prior associated with the states to be distinguished is assumed. 
%used and \cite{bkm2011} the author described another approach based on isoenergetic compression, which also enables under special conditions a break of the well-known Hailstorm bound.

 Two disparate concepts are combined in the paper. These are quantum state discrimination and the modification of the  quantum potential resulting in a transformation of the wave function. % associated with the insertion of a barrier.
 Bayesian hypothesis testing, the particular form of quantum state discrimination investigated here, was developed by Helstrom and others \cite{helstrom, holevo, yuen}. For a recent paper on the topic of state discrimination  between two possible states with given prior and transition probability see Brody {\it et al.} \cite{dbbm}. It is generally accepted, but will be challenged in the paper, that the optimal Bayes cost in the binary case, is given by the Helstrom bound, which only depends on the prior and the transition probability between the states and can be written in a simple closed form. 
%We will show that under some special conditions, including the addition of energy, there is a measurement strategy that improves on the Helstrom bound.
The second concept is the ability to modify wave functions  in a beneficial way, if one inserts impenetrable barriers corresponding to a potential spike in a simple configuration space, here chosen to be a ring.
%%%%%%%%%%%%
%%%%%%%%%%This paper is part of a series \cite{bkm2011, bkm2011a, bkm2010} analysing the Helstrom bound.
%%%%%%%%%%%%%%%%%%
   In earlier papers \cite{bkm2011, bkm2011a} these two ideas were applied to study the insertion of one barrier into a ring or a one-dimensional infinite square well both adiabatically and instantaneously. The current approach is simpler by exploiting  the existing nodes of the candidate wave functions and by focusing on the link between the energy required to insert the barrier and the candidate wave functions.

   %to the rotational symmetry of the ring, which is only broken once the barrier is inserted, and the eigenvalue degeneracy of the Hamiltonian. 
 % of energy conservation and transitions between
%some of which  will be reviewed and exploited in this paper.
%These insights are combined with the analysis of As a result one can lower the Bayes cost, also called the minimum error probability, in special cases below the traditional optimal bound.

 %In the instantaneous case the transition probability between the states stayed unchanged, but the insertion energy is different for the two candidate states. In the adiabatic case no energy is required, but the resulting candidate wave functions have a lower transition probability. 

Next, a description of the setup and the procedure to be analysed. %We discuss thdistinguish  the instantaneous and the adiabatic case.  These are distinguished by the configuration space used, either the  wave function is spread over one or two rings, and by the barrier insertion speed.
%A candidate wave function is defined on a ring.
 The following decision problem is presented. With equal probability, prior of $1/2$,
one of two quantum states is put into the configuration space - a ring. Our challenge is to determine with the smallest possible error, which state has been selected. 
Two strategies for calculating the Bayes cost are proposed. In the first strategy,  the combination of prior and transition probability between the two quantum states alone is sufficient to calculate the conventional optimal minimum error probability, i.e. the Helstrom bound, prior to the insertion of any barrier. %This
%error probability can be achieved by a well-known measurement strategy \cite{helstrom} not further discussed in the paper.
The standard procedure, reliant on the optimal POVM as described in the book by Helstrom\cite{helstrom}, results in the following binary decision   cost 
\begin{eqnarray}
%C(|\bra\phi|\psi\ket|^2)%\nonumber\\
%&=&\frac{1}{2}-\frac{1}{2} \sqrt{1-|\bra\phi_{before}|\chi_{before}\ket|^2},\nonumber
\frac{1}{2}-\frac{1}{2}\sqrt{1-\cos^2(\alpha)},\nonumber
\end{eqnarray}
where $\cos^2(\alpha)$ is the transition probability between the two candidate states, and cost $1$ is assigned to an incorrect  and cost $0$ for a correct decision.

  The second  strategy for calculating the error probability is novel. The wave function is first modified by the simultaneous insertion of two barriers  breaking the symmetry of the configuration space.  %Therefore, we first break the   symmetry of the configuration space (either one or two rings) by barrier insertion. 
  The insertion of the barrier can be carried out with different speeds. We consider an extreme case: instantaneous insertion. This insertion can require energy and can modify the wave function, since its amplitude at the impenetrable barrier location will be zero and the expansion in the new bases to reproduce the original amplitude, except at the insertion points, is accompanied with a change of energy. The modified candidate wave function, now entangled with the barriers, is  probed and the  new binary decision cost estimated. It is shown that the extended wave functions incorporating the barriers can be orthogonal, even if the original overlap was non-zero.%or the change in the transition probability of the modified candidate states (in the adiabatic case) is noticed. In both situations the Helstrom bound can be broken.

%With equal probabilities (prior: $\xi=1/2$) one of two quantum states is put on a ring in the instantaneous case. The symmetry of the ring is then broken by the instantaneous insertion of a barrier.
%In the adiabatic case studied subsequently the candidate wave function are spread out on two  rings and barriers are simultaneously  inserted in both rings. In both cases the insertion of a barrier modifies the wave function.

  %To keep the energy of the system unchanged, the probability weights associated with the changing eigenfunctions are dynamically modified. The reduced chamber is then probed using conventional measurement strategies and the Bayes cost calculated, which in some cases can be below the Helstrom bound. Entropy maximization associated  with the isoenergtic compression leads to an non-linearity and is the key to understand the result.
%For all specific quantum states the Helstrom bound is always  higher or equal than the new Bayes cost.

%In the appendix we consider what happens, if instead of enlarging the chamber we insert a impenetrable barrier at a non-nodal  point inside the original chamber. %subsequent section we consider the instantaneous expansion of the chamber.

%In the present paper we study the Bayesian approach to a binary decision problem for a particles in a box.

%The experimental set up considered is a
%We will show that the combination of these two concepts  allows a novel evaluation of the Bayes cost.
%This is done by evaluating the different types of Bayes cost before and after the insertion of the barrier.

Two motivations for this research stand out. On the one hand it sheds some light on foundational issues in quantum measurement theory, and on the other hand  a plethora of problems in quantum information theory depend on optimal state discrimination.

Quantum mechanics on the ring is specified by the Hamiltonian and boundary conditions.
%includes the following properties. 
%Some useful properties of quantum mechanics on a ring
% Some basic facts about a particle in a periodic potential are described.
%The particle in a periodic potential In the following section first the instantaneous and then the adiabatic insertion of impenetrable barriers into an infinite circular square well at both nodal or non-nodal points is presented.
%The set up is
%a wave function $\chi(x)$  of
%a particle of mass $M$  trapped in a one-dimensional infinite square well of width $L$. The Hamiltonian is given by
%\begin{eqnarray}
%H= - \frac{\hbar^2}{2M}\frac{d^2}{dx^2}.
%\end{eqnarray}
 The Hamiltonian of a particle trapped  on a ring of radius one  is 
\begin{eqnarray}
H= - \frac{\hbar^2}{2M }\frac{d^2}{d\theta^2}\nonumber
\end{eqnarray}
 with energy eigenvalues $E_n=\frac{  \hbar^2}{2 M}n^2$ for $n\in \mathbb{N}$.
The wave function is defined for $\theta\in[0,2 \pi] $.

 %study implications for the distinguishability of states, if the insertion is carried out on a circle.
 
% The eigenfunctions of the Hamiltonian  are 
 %\begin{eqnarray}
%\phi_n^{\pm} (\theta)= \frac{1}{\sqrt{\pi}}e^{\pm\imath n\theta}.\nonumber
%\end{eqnarray}
%and
%\begin{eqnarray}
%\psi_n (x)= \sqrt{ \frac{2}{L}}\cos\Big(\frac{n \pi x }{L}\Big),\nonumber
%\end{eqnarray}
%with the respective energy eigenvalues $E_n(L)=\frac{\pi^2 \hbar^2 n^2}{2 M R^2}$.
%Here we expand instantaneously the compartment by  an amount $\delta$.

%In the later part of the paper we calculate for ...

The structure of the rest of paper is as follows. In section II   the impact of an instantaneous insertion of a barrier on a ring is studied.  In section III the simultaneous insertion of two barriers is considered. The  binary choice problem  between two quantum states is  tackled at the end of the section. 
 %how a negative measurement can be used to change the angle between the states simultaneously with the weight in such as to leave the Bayes cost unchanged.
%In the penultimate section progress is made in establishing a new bound for the Bayes cost.
In the conclusion the result is briefly reviewed  and some general comments added.
%In the appendix  the insertion of an impenetrable barrier is described and the cost calculated.

%In an appendix it is shown what kind of measurements leave the Helstrom bound unchanged.

 %to gain a better understanding of the importance of measurement in contrast to information erasure on the Maxwell daemon. Only the combination of measurement and information erasure, which are in some ways complementary, are sufficient to explain Maxwell's daemon \cite{SagawaUeda09}.
% It was discovered that the simple insertion of a barrier itself at a non-nodal point is associated with work and %changes the energy of the system by creating fractal wave function \cite{berry}. Th

%Here ideas from an earlier paper about minimum decision cost for distinguishing states in the Bayesian framework %\cite{bm96} are combined with a more recent paper on Maxwell's demon \cite{bbm} to obtain some new bounds for quantum %state discrimination.

%Optimal state discrimination is a prominent problem in quantum information theory,

\vspace{-.498cm}
 \section{Instantaneous insertion of  a barrier }
\label{sec:1b}
\vspace{-.398cm}
\noindent
In this section the barrier insertion, considered to be instantaneous, at both nodal and non-nodal points is reviewed. 
%In this section the instantaneous insertion of impenetrable
%barriers at both nodal or non-nodal points is presented
%In the next subsection an instantaneous insertion of the barrier at the origin is considered.
%In the following subsections  the insertion of a barrier adiabatically and instantaneously is discussed 
%, which  can be divided into four cases. In the first two cases one inserts the barrier at
%a nodal point, and in the second case the insertion happens at a point of non-zero amplitude.
%In the first case the calculation is  easier, since the energy of the system is left unchanged.
%Let us look at the example of the second excited state
%\begin{eqnarray}
%\phi_{before}(x)=\sqrt{\frac{2}{L}}\sin\Big(\frac{2 \pi x} {L}\Big).\nonumber
%\end{eqnarray}
The nodal point insertion is dealt with first. This is easier,
since the particle wave function and energy is left unchanged -  for background material see section II of  Bender {\it et al.} \cite{bbm}, where %This means the experimenter can carry out the procedure
%without adding or extracting energy.  
 a series of results for a particle in a one dimensional box, directly applicable to quantum mechanics on a ring, were established.
%For more details see section VII of  Bender {\it et al.} \cite{bbm} and and section II of \cite{bkm2011}. 
%Barrier and wave function on the ring do not get entangled, but instead form a product state.
As an aside, we only call a point a node, or more correctly a `fixed node', if the amplitudes  at this point stays zero at all times. 
Wave functions that are superposition of eigenfunctions of $H$ can have zero amplitude points that change with time. These we do not consider, since an insertion at a `transitory node' can require energy.

%Significantly, the expectation value of the Hamiltonian
%before and after the insertion remains unchanged, since
%in the definition of the energy both the size of the well
%and the counter of the energy level appear as squares.
%This means the experimenter can carry out the procedure
%without adding or extracting energy
 
%If the insertion happens at a nodal point, where the amplitude  vanishes, the wave function is unaffected and no energy is needed  to insert the barrier, see section VII of Bender {\it et al.} \cite{bbm}. 

The situation is  more intricate  for an insertion at a non-nodal point.
Energy is needed to modify the wave function. %, if the insertion happens at a point, where the amplitude is non-vanishing.
 In the idealised setting considered here the required energy  is infinite. The energy localised in
the barrier point inserted at $t=0$ propagates through the system at $t > 0$
and increases the energy on the ring. The result
is a fractal wave function. The details of the calculation can be found in sections  IV \& VI of Bender {\it et al.} \cite{bbm} or in section II of \cite{bkm2011} \& \cite{bkm2011a}.

% Based on the argument fully laid out in section IV of \cite{bbm}, one can say that the Fourier expansion converges  pointwise everywhere except at the insertion point, i.e. the  Gibbs phenomenon.
%$t=0$ at the point $\theta=0$ into  $\phi$

%To summarise,
%carried out in section 4 of the earlier cited paper by Bender  {\it et al.} \cite{bbm}, which forms the backbone of this section,
%the energy needed to insert the barrier instantaneously at a non-nodal point is infinite. 

%The insertion of the barrier at nodal and non-nodal points was considered for infinite potential well in section IV  \cite{bbm}. The analysis here is very similar.

\vspace{-.498cm}
\section{The instantaneous insertion of two impenetrable barriers on a ring}
\vspace{-.398cm}

%After the instantaneous insertion discussed above, we turn to the other extreme:  adiabatic insertion.
\noindent
The case of two simultaneous insertions changing the ring into two separate infinite square wells is considered in this section. 
%Like the previous case the insertion of two barriers represents an idealisation, since the insertions are not only assumed to happen instantaneously but also simultaneously.  %The process is assume to happen infinitely fast. 
In the following paragraphs the cost is evaluated before and after  the insertion of two barriers for distinguishing the following two candidate wave functions defined, at $t=0$, as 
\begin{eqnarray}
\phi(\theta)&:= &\frac{1}{\sqrt{\pi}}\sin(  \theta ),\nonumber\\
\psi(\theta)&:=&  \frac{1}{\sqrt{\pi}}\sin(\theta - \alpha),\nonumber
\end{eqnarray}  
where $\theta \in [0, 2 \pi]$, and $\alpha\in(0,\pi/2)$. 
The initial overlap of the wave functions is
 \begin{eqnarray}
| \bra \phi | \psi \ket |^2= 
\frac{1}{\pi^2} \Big|\int_{0}^{2\pi} d \theta \sin ( \theta )\sin( \theta + \alpha)\Big|^2 =\cos^2(\alpha),\nonumber
\end{eqnarray}
and the standard Helstrom cost is given by 
\begin{eqnarray}
%C(|\bra\phi|\psi\ket|^2)%\nonumber\\
%&=&\frac{1}{2}-\frac{1}{2} \sqrt{1-|\bra\phi_{before}|\chi_{before}\ket|^2},\nonumber
\frac{1}{2}-\frac{1}{2}\sqrt{1-\cos^2(\alpha)}.\nonumber
\end{eqnarray}

Due to the pre-insertion symmetry  both candidate wave functions are eigenfunctions of the Hamiltonian of a particle on a ring. The symmetry is only broken by  the barriers. The barriers are inserted at the point $0$ and $\alpha$ at time $t=0$. 
The barrier inserted at point $0$   leaves $\phi$ unchanged, but the second barrier at point $\alpha$ hits a non-nodal point.
The situation is the reverse for $\psi$, since it lacks a node at $\alpha$, but has a node at $0$. As explained before, at nodal points no energy transfer occurs, but a barrier at a non-nodal point is associated with a change of energy.

An instantaneous insertion requires, due to the change of the configuration space,  an expansion of the original wave functions into the energy basis of the two separate one dimensional infinite wells. This will be carried out next.
The first expansion is in the interval $(0,\alpha)$ with the discrete energy levels $E^{\alpha}_n=\frac{  n^2 \pi^2 \hbar^2}{2 M \alpha^2}$  and the second expansion is for the interval  $(\alpha,2 \pi)$ with the discrete energy levels $E^{2\pi-\alpha}_n=\frac{  n^2 \pi^2\hbar^2}{2 M (2\pi-\alpha)^2}$ such that the first candidate wave function has directly after the insertion %, at $t=0^+$, 
the following form 
\begin{eqnarray}
\phi_{after}(\theta):= 
\left\{
\begin{array}{lr}
\sqrt{\frac{1}{\pi}}\sum_{n=1}^{\infty}a_n \sin\Big( \frac{n \pi\theta} { \alpha}\Big)  &0< \theta < \alpha,\nonumber\\
\sqrt{\frac{1}{\pi}}\sum_{n=1}^{\infty}b_n\sin\Big( \frac{n  \pi\theta }{2\pi-\alpha} \Big) & \alpha< \theta < 2\pi,\\
\end{array} 
 \right. \nonumber
\end{eqnarray}
where 
\begin{eqnarray}
a_n&:=&  \frac{1}{\pi}\int_{0}^{\alpha} d\theta  \sin(\theta) \sin\Big( \frac{n \pi\theta} { \alpha}\Big)  \nonumber\\
&=& (-1)^n\ \frac{\alpha \, n }{\alpha^2-\pi^2 n^2} \sin(\alpha),\nonumber
\end{eqnarray}
and 
\begin{eqnarray}
b_n&:=&  \frac{1}{\pi}\int_{a}^{2\pi} d\theta  \sin(\theta)\sin\Big( \frac{n  \pi\theta }{2\pi-\alpha} \Big) \nonumber\\
&=& -\frac{(2\pi-\alpha)n}{(\alpha-(n+2)\pi)(\alpha+(n-2)\pi)}\sin(a).\nonumber
\end{eqnarray}
Similarly, the transition probability of $\phi$ to the combination of the eigenfunctions of energy $E^{\alpha}_n$ and $E^{2\pi-\alpha}_m$
  is $|a_n b_m|^2$, and the energy transfer from the barrier inserted at the non-nodal point $\alpha$ is 
 \begin{eqnarray}
 \Delta E^{\phi}_{nm}:= % \frac{ n^2 \pi^2\hbar^2}{2 M \alpha^2} +\frac{  m^2 \pi^2\hbar^2}{2 M (2\pi-\alpha)^2}-\frac{ \hbar^2 }{ 2M }\nonumber\\
  \frac{  \pi^2\hbar^2}{2 M }\Big(\frac{n^2}{\alpha^2}+\frac{m^2}{(2\pi- \alpha)^2}-\frac{1}{4 \pi^2}\Big).\nonumber 
 %&=& \frac{ \hbar^2 \pi^2}{2 M }\Big(\frac{n^2(2\pi- a)^2\pi^2+m^2a^2\pi^2-(2\pi- a)^2 a^2}{a^2(2\pi- a)^2\pi^2}\Big)\nonumber\\
%&=&  \frac{ \hbar^2 \pi^2}{2 M }\Big(\frac{(n^2+m^2) a^2 \pi^2 +4n^2\pi^3(\pi-a)-4\pi^3 a}{a^2(2\pi- a)^2\pi^2}\Big)\nonumber
 \end{eqnarray}
 By an appropriate choice of $\alpha$, e.g. $\alpha=\pi/4$,  the energy change $\Delta E^{\phi}_{nm}$ is always positive. 
  Similary, the second candidate wave function can be expanded into
\begin{eqnarray}
\psi_{after}(\theta):=
\left\{
\begin{array}{lr}
\sqrt{\frac{1}{\pi}}\sum_{n=1}^{\infty}c_n \sin\Big( \frac{n \pi \theta} { \alpha}\Big)  &0<\theta <\alpha,\nonumber\\
\sqrt{\frac{1}{\pi}}\sum_{n=1}^{\infty}d_n\sin\Big( \frac{n  \pi\theta }{2\pi-\alpha} \Big) & \alpha< \theta<  2\pi,\nonumber
\end{array} 
 \right. \nonumber
\end{eqnarray}
where 
\begin{eqnarray}
c_n&:=&  \frac{1}{\pi}\int_{0}^{\alpha} d\theta  \sin(\theta-\alpha) \sin\Big( \frac{n\pi\theta} { \alpha}\Big)   \nonumber\\
&=& \frac{\alpha \, n  }{\alpha^2-\pi^2 n^2}\sin(\alpha)\nonumber
\end{eqnarray}
and 
\begin{eqnarray}
d_n&:=&  \frac{1}{\pi}\int_{\alpha}^{2\pi} d\theta  \sin(\theta-\alpha) \sin\Big( \frac{n  \pi\theta }{2\pi-\alpha} \Big) \nonumber\\
&=& (-1)^n  \frac{(2\pi-\alpha)n}{(\alpha-(n+2)\pi)(\alpha+(n-2)\pi)}\sin(\alpha).\nonumber
\end{eqnarray}
The transition probability of $\psi$ to the combination of the eigenfunctions of $E^{a}_n$ and $E^{2\pi-a}_m$
is $|c_n d_m|^2$, and the energy transfer from the barrier inserted at the non-nodal point $0$ is again $\Delta E^{\psi}_{nm}$.
%= \frac{  n^2 \pi^2 \hbar^2}{2 M \alpha^2} +\frac{  m^2 \pi^2\hbar^2}{2 M (2\pi-\alpha)^2}-\frac{ \hbar^2 }{ 2M }$. As before for an appropriate choice of $\alpha$,   the energy change $\Delta E^{nm}_{\phi}$ is always positive. 

Different elements in the expansion  are associated to different energies. 
Due to energy conservation there has to be a source for this change of energy, which will be always positive for a value of $\alpha$ of $\pi/4$. This additional energy can only come from the barrier inserted at the non-nodal point. A laser beam could be a possible realisation of the barrier. 
%This energy is transferred to the wave function during the insertion from the barrier, possibly realised through a laser beam. 
The photons of the laser beam, or any other realisation of the barrier, have an energy dependent entanglement with the expanded wave function on the ring, where each combination of energy levels 
in the two chambers is associated with a complementary state for the barriers to achieve energy conservation.
% The nodes of the two candidate wave functions are at
%different points and correspond to separate barrier insertion points. 
The energy transfer is  to $\phi$ from the  barrier at $\alpha$ and to $\psi$ from the barrier at $0$, since each candidate wave function has its energy only modified through one specific barrier corresponding to the initial non-nodal point.

The extended wave function including the barrier can be written before the insertion as either
\begin{eqnarray}
\Phi_{before}= \psi \otimes \omega^{before}_0(0) \otimes  \omega^{before}_0(\alpha),\nonumber
\end{eqnarray}
or
\begin{eqnarray}
\Psi_{before}= \phi \otimes \omega^{before}_0(0) \otimes  \omega^{before}_0 (\alpha),\nonumber
\end{eqnarray}
where $ \omega^{before}_0(0)$ and $\omega^{before}_0(\alpha)$ correspond to the wave functions associated with the pre-insertion 
barriers at the points $0$ and $\alpha$ respectively.
Directly after the insertion %, at $t=0^+$, 
the  extended wave functions are transformed into
\begin{eqnarray}
\Phi_{after}= \sum_{n,m=1}^{\infty} a_n b_m \sin\Big( \frac{n \pi x} { \alpha}\Big)\nonumber\\
 \otimes   \sin\Big( \frac{m \pi y}  { 2\pi-\alpha}\Big) \otimes 
\omega^{after}_{0}(0) \otimes \omega^{after}_{n,m}(\alpha)\nonumber 
\end{eqnarray}
and 
\begin{eqnarray}
\Psi_{after}= \sum_{n,m=1}^{\infty} c_n  d_m \sin\Big( \frac{n \pi  x} { \alpha}\Big) \nonumber\\
\otimes  \sin\Big( \frac{m \pi y}  { 2\pi-\alpha}\Big) \otimes 
\omega^{after}_{n,m}(0) \otimes \omega^{after}_{0}(\alpha)\nonumber 
\end{eqnarray}
where $x\in(0,\alpha)$ and $y\in(\alpha, 2\pi)$, and  
$\omega^{after}_0(0)$ $\&$ $\omega^{after}_0(\alpha)$ are the  barrier  wave functions, if  inserted at either a nodal point at $0$ or $\alpha$. 
$\omega^{after}_{n,m}(\alpha)$ and $\omega^{after}_{n,m}(0)$ correspond to barriers that transferred $\Delta E_{nm}$ to the candidate wave functions $\Phi_{after}$ and $\Psi_{after}$ respectively. 
%\begin{eqnarray}
 %\Delta E_{nm} =\frac{  \pi^2\hbar^2}{2 M }\Big(\frac{n^2}{\alpha^2}+\frac{m^2}{(2\pi- \alpha)^2}-\frac{1}{\pi^2}\Big).\nonumber 
 %&=& \frac{ \hbar^2 \pi^2}{2 M }\Big(\frac{n^2(2\pi- a)^2\pi^2+m^2a^2\pi^2-(2\pi- a)^2 a^2}{a^2(2\pi- a)^2\pi^2}\Big)\nonumber\\
%&=&  \frac{ \hbar^2 \pi^2}{2 M }\Big(\frac{(n^2+m^2) a^2 \pi^2 +4n^2\pi^3(\pi-a)-4\pi^3 a}{a^2(2\pi- a)^2\pi^2}\Big)\nonumber
 %\end{eqnarray}
% which is always  positive for $\alpha=\pi/4$.
The transition probability  for $\alpha$ taken to be $\pi/4$  is therefore altered during the insertion processes and has the form %and stays unchanged 
\begin{eqnarray}
&&
| \bra \Phi_{after} | \Psi_{after} \ket |^2 \nonumber\\
&=&
 |\sum_{n,m=1}^{\infty} a_n c_n b_m d_m  \bra \omega^{after}_0 (0) |\omega^{after}_{nm} (0) \ket \nonumber\\
 &&
\bra   \omega^{after}_{nm} (\pi/4) |\omega^{after}_0 (\pi/4) \ket|^2=0,\nonumber
\end{eqnarray}
because both $ \bra \omega^{after}_0 (0) |\omega^{after}_{nm} (0) \ket$ and $\bra   \omega^{after}_{nm} (\pi/4) |\omega^{after}_0 (\pi/4) \ket$
are zero for all $n$ and $m$. It follows from the association of the different barrier states, e.g. photon states representing a laser beam, with different energies. % , e.g. in the case of a laser beam some of the photons in the product state will have reduced energies. %, since wave-functions associated with different energy are orthogonal.

 After the insertion, the entanglement of the two candidate wave functions with the respective barriers results in two states, which are orthogonal. As a consequence, as an improvement from the conventional Helstrom bound, the associated binary decision cost reduces to zero. 
In the upcoming conclusion the result is reviewed and some comments added.

%Any pair of initial states with overlap $1/2$ can be transformed prior to the isoenergetic compression into states with large energy differences without effecting the distance between the states.
% The wave function $\phi_{before}$ as a result of the compression, if $\delta$ is chosen as before, yields the new ground state.
%
%\begin{eqnarray}
%\frac{1}{2}
%\Big(\bra 1|M^\dag M|1\ket+ \bra 2|M^\dag M|2\ket+ 2\bra 1|M^\dag M|2\ket\Big)+O(\epsilon^3)=
%\frac{1}{2}\Big(2-2 \frac{40}{9}\epsilon^2\Big)+O(\epsilon^3).\nonumber
%\end{eqnarray}
%It was shown in the previous section that for the second test function $\psi(x)$ the weight in $|1'\ket$ can be set equal to $\epsilon$, where $\epsilon$ is a function of $N$. Further we showed that $\epsilon$ will become smaller than any fixed number, if one chooses $N$ large enough.
%The inner product of $\phi$ and $\psi$ after the compression is
%\begin{eqnarray}
%|\bra\phi_{after} | \psi_{after} \ket|^2=\epsilon.\nonumber
%\end{eqnarray}
%where $\epsilon$ is a function of $|N\ket$ and can be made arbitrarily small.
%The new inner product between the states can therefore be reduced,
%and as a result the new cost
%\begin{eqnarray}
%\frac{1}{2}-\frac{1}{2}\sqrt{1-4\xi(1-\xi) \epsilon}\nonumber
%\end{eqnarray}
%can be lower than the Helstrom bound associated with the overlap $\alpha$ before the compression and can even be arbitrarily close to $0$.
%The probability for the particle to be in the $[0,L-\delta]$-interval is
%\begin{eqnarray}
%Prob_{L}=1-X .\nonumber
%\end{eqnarray}
\vspace{-.698cm}
\section{Conclusion}
\vspace{-.398cm}
\noindent
The aim of the paper was to show that the Helstrom bound in the binary quantum discrimination setting can be breached.
Inserting a barrier instantaneously in a ring at a non-nodal point always requires energy, whereas an insertion at a nodal point leaves the energy unchanged. 
If two barriers are inserted, one at a node and one at a non-nodal point, then there is only energy-level dependent entanglement of the wave function on the ring    with the barrier at the non-nodal point. % and influenced the transition probability of the states. Each element of a superposition of energy eigenstates gained during the barrier insertion  a different amount of energy. %This entanglement reduced the overlap between the states.% This avoids having to resort to and consequently energy-level dependent entanglement with the barrier. 
%Here we dealt with the problem
Furthermore, by having the node of one candidate wave function to be the non-nodal point of the other, one can ensure
that the energy transfer in the two cases is to barriers at distinct points.
%two candidate wave functions, where the node of the energy transfer from the different candidate wave functions to different barriers.%, since out of the two barriers one is always inserted at a node, and the node of one candidate wave function is the non-nodal point of the other.  
%This leads to an energy-dependent entanglement between candidate wave functions and different barriers 
This can be used to help distinguish between different states. 
 An extension beyond an instantaneous insertion should be possible, since any non-adiabatic insertion at a non-nodal point needs energy. 
Nevertheless,  a careful analysis of the finite speed case is required.
% A critical point, which hampered the analysis of the particle in a box in an earlier paper \cite{bkm2011}, was the reliance on superpositions of eigenfunctions. %In the adiabatic case the minimal energy transformation associated with the insertion at a non-nodal point leads to a shift of the wave function changing the transition probability between candidate states. In both cases the binary  cost can be lowered below the Helstrom bound.
%An insertion with a speed between the two extremes, not explicitly discussed, should be handled in a similar way, since the finite speed insertion also requires energy similar to the instantaneous insertion.  
The exact cost reduction depends on how precisely one can place the barrier, the width of the barrier, the speed of insertion, i.e. three issues related to time evolution of the potential associated with the barrier.
%, and the measurement uncertainty associated with the energy needed to insert the barrier.
If each of these points can be addressed satisfactorily, then not only in the idealised situation can one obtain an improvement of the optimal cost beyond the  Helstrom bound.

%Two potentially controversial assumption underpin this process.
 %The first assumption is tied to the insertion of an impenetrable barrier of infinitesimal width.  
 %This type of procedure was already discussed in different forms, i.e. from the instantaneous to the adiabatic, in an earlier paper by
%Bender {\it et al.}  \cite{bbm}. 
%It is an idealisation, but still has some experimental relevance, since % an insertion with finite speed also requires energy. 
% a slower, but not infinitely slow, insertion of a barrier also requires a finite amount of work. The finite speed case is harder to analyse and  demands the solution of a time-dependent Hamiltonian. %An insertion at any finite speed would give a clue about the state one is probing.
%The only main difference is the value of the energy required and the increased difficulty to carry out the more realistic calculation.
%The result described in the previous paragraphs  can in a technically more involved way also be derived using an isothermal contraction.
%The second assumption is that one is able to measure the energy needed to insert a barrier.  A fuller analysis requires a realistic interpretation of what    it means to insert a barrier, e.g. change the potential with time or let a laser beam interact with the test particle, and how the barrier interacts with the wave function. %The two test states, which initially had an overlap of $1/2$, have after the contraction a reduced overlap.

Naturally, one can criticise the failure to provide a realistic time evolution; %of the barrier 
%as well as the lack of an explicit calculation of the time evolution of the  state
 only the infinitely  fast insertion was examined. In defence, one can point to the idealised nature of the proposal and the statement that more realistic examples can be viewed as an extrapolation of the procedure under consideration. 

The following Gedankenexperiment might be instructive, because  it shows that it is possible to  construct an example were no information is transferred to the experimenter, when the potential representing the barrier is altered. 
Imagine an experimenter doing a fixed amount of work per unit of time to insert the barrier, i.e. pushes in the barrier with constant power.  % then  no information is transferred to the experimenter, since the power required is independent of the test function, when the  potential representing the wall is changed. 
   Dependent on the test wave function the change of the potential is either larger or smaller. The potential takes on different shapes and
 affect the states in different ways, ergo can change the distance between the states, without direct, energy based, leakage of information to the outside.
% This is achieved by letting the experimenter do a fixed amount of work per unit of time. Dependent on the test function the change of the potential is either larger
%not been studied, but should be of interest.

 %As an aside,  the instantaneous case can be extended trivially to the two ring configuration, and any possible two ring configuration can be mapped into a one ring configuration. 

%An intuitive way to grasp the result is to link the additional information provided by the energy required to insert the barrier, i.e.
%the insertion of a barrier into non-nodal is more energy intensive than the insertion at a nodal point, to the ability to distinguish the states.

%One could also consider the instantaneous insertion of a barrier into a well, but in this case the energy required would be for non-nodal points normally infinite, which is unrealistic.

Questions about the possibility of superluminal communication can be raised, but one should keep in mind that the Schr\"{o}dinger equation, a diffusion equation without a propagation speed limit, has its limitations. % as well as the fact that the Schroedinger equation, since it is a diffusion equation, does not posses an upper limit for the speed with which information travels, and therefore standard non-relativistic quantum mechanics .

 %As a caveat, in the current derivation breaching the bound is only possible, if one pumps  an infinite amount of energy into the system through the instantaneous expansion of the square well.
% Maybe it is worth delving into the question, why it is possible to break the Helstrom bound? One explanation could come from the added degrees of freedom accessible through the insertion of a barrier at an appropriate position.  Similarities exist to an earlier paper by Bender {\it et al.} \cite{bbm}, where the insertion of a barrier led to an interesting entropy effect.
%A more general lower bound will be discussed in a companion paper.
%Two motivations for the work presented immediately spring  to mind. Light is shed on foundational issues in quantum measurement theory, and a plethora of problems in quantum information theory depend on optimal state discrimination, since 
%almost without fail quantum algorithms can be viewed as procedures to distinguish between states.
Almost without fail quantum algorithms can be viewed as procedures to distinguish between different states.  The method described above works for states with arbitrary overlap and should find application in the area of quantum algorithm. 
%This is  to be examined in a separate paper. 
 %The important step is always to map the two or more candidate wave functions with arbitrary overlap into
  %states with distinct nodal points without initially changing the distance between the candidate states.

 Experimental implementation, for example in the area of Bose-Einstein condensate or ion traps,   with a laser beam as a barrier, should be of interest. 
% The invariance of the Helstrom bound to operations that leave the product of the priors and the square of the transition probability unchanged will be exploited.
%The improvement in the cost beyond the Helstrom bound through the division of the box can be explicitly calculated.
%Possible implications of idea for quantum information theory will also be discussed in the following paper.

\hspace{-.38cm}The author wishes to express his gratitude to D.C. Brody for  stimulating discussions.
%Many implications of the described method are quite obvious and will only be briefly noted here.
%1) As one can distinguish states more ....
%What is the optimal procedure to separate two states is not the subject of the paper, instead it was to show that the common procedure are too limited to cover the full subtly of quantum mechanics.

%On a more personal note, it has been the third attempt by the author to write something meaningful about quantum algorithm.
%

%\twocolumn
%\end{multicols}
 
   %\begin{multicols}{1}%[\textbf{Example for a three column text}]
    % \blindtext

\vspace{-.38cm}
 %  \end{multicols}

\newpage
\appendix
\onecolumngrid
%\appendix
\vspace{-.28cm}
\section{ An Example involving an intermediate measurement}
\vspace{-.198cm}
\noindent 
In the appendix,  the zero-one cost %for  state discrimination
 %involving an intermediate measurement %between two candidate states
 is determined for a discrimnation strategy that %reproduces the above result but 
employs as a first step a POVM to update the input state. The probability weighted  output of this process is then %measurement
%The second strategy for calculating the error probability is novel. An intermediate POVM is first applied
  probed using the conventional Helstrom method.
 % The %resulting 
%  output is then
%candidate density matrices 
%  analysed
%using the conventional Helstrom strategy. 
The combined new error probability is calculated. It can break the straightforward Helstrom bound, which only depends on the prior and the transition probability,    if  the candidate states  and the components of the intermediate POVM are judiciously chosen. 
% `negative measurement'\cite{renn, renn1}. %or the L\"uders Postulate. 
%After the initial updating of the state %, which produces a `positive' result with a small probability 
%the conventional Helstrom approach is employed. %update by  %for probed via an intermediate POVM. 
%In the appendix, binary state discrimination minimising the expected zero-one cost %, where cost one is assigned to an incorrect and cost zero for a correct decision, 
%is again the topic under investigation. %The existence of a prior associated with the states to be distinguished is assumed.
%One of two quantum states is sent to an experimenter. The challenge is to determine with the smallest possible error, which state has been selected. 
%This novel way to calculate the Bayes cost is compared  to the conventional optimal minimum error probability, i.e. the Helstrom bound, which only depends on the prior and the transition probability. %. In the first strategy, the combination of prior and transition probability between the two quantum states alone is sufficient to calculate 
%This is done via the insertion of 
%an intermediate N-component positive operator-valued measure (POVM), the candidate state is  measured and a choice is made. %The associated cost for this procedure is calculated.%The aim is to minimze the zero-one discrimination cost. % in the conventional way. %to update  the candidate state before the final analysis. %through the application of a N-component POVM.
The measurement induced updating of the density matrix involves a normalization\footnote{The normalisation adds non-linearity to the process, since  $  \frac{\pi^\dagger( \rho +\sigma) \pi}{{\rm Tr}(\pi^\dagger(\sigma+\rho) \pi)} $ 
is not necessarily equal to the probability weighted sum of    $\frac{\pi^\dagger\rho \pi}{ {\rm Tr}(\pi^\dagger \rho \pi) } $ and $
 \frac{\pi^\dagger\sigma \pi}{{\rm Tr}(\pi^\dagger\sigma \pi)} $.}
  and gives the process a non-linear twist. 
  
%This reproduces the result of the earlier part of the  paper without relying on the L\"uders Postulate and without requiring excessive precision in implementation. 
%\subsection{The POVM associated intermediary m
Notation is next introduced conducive to the problem at hand. %, i.e.$\cos$ and $\sin$ are replaced by the letters $c$ and $s$ respectively. % The two selected states are now called $\sigma$ and $\rho$ of the form . 
The candidate states are  $|\psi\ket=\sin{\alpha}|0\ket +\cos{\alpha}|1\ket $ and  $|\phi\ket=\sin{\beta} |0\ket + \cos{\beta}|1\ket $, more conveniently written as $|\phi\ket = \cos{\delta} |\psi\ket  + \sin{\delta} |\psi_{\bot}\ket  $  with \begin{eqnarray}
 \cos^2{\delta}=( \sin{\alpha} \sin{\beta} + \cos{\alpha} \cos{\beta})^2= \cos^2(\alpha-\beta).\nonumber
\end{eqnarray}  The  density matrices are $\rho= |\phi\ket \bra \phi | $  and $\sigma= |\psi\ket \bra \psi | $  with the respective priors  of $\xi$ and $1-\xi$. 
%The intermediate POVM has the form.
The `in-between measurement' has a POVM associated with it, which
%acting on the candidate state 
is a resolution of the identity of the form $\mathbf{I}= \sum_{i=1}^N \mathbf{\Pi}_i$, with the $\mathbf{\Pi}_i$ being positive semi-definite matrices. %Hermitian matrices with non-negative eigenvalues. %+ \mathbf{\Pi_2}+\mathbf{\Pi_3}$,
%The  Hermitian matrices
%  \begin{eqnarray}
%\mathbf{\Pi_1}= \left(
%\begin{array}{cc}
%\lambda_1^2 & 0 \\
%0 & \lambda_2^2\\
%\end{array}\right), \,\,\,
%\mathbf{\Pi_2}= \left(
%\begin{array}{cc}
%(1-\lambda_1^2)/2 & \zeta \\
%\zeta^* & (1-\lambda_2^2)/2\\
%\end{array}\right),  \,\,\, \nonumber%
%%\right)
%\mathbf{\Pi_3}= \left(
%\begin{array}{cc}
%(1-\lambda_1^2)/2 & -\zeta \\
%-\zeta* & (1-\lambda_2^2)/2\\
%\end{array}\right), \nonumber
%\end{eqnarray}
%with $\zeta$ chosen so that the eigenstates of $\mathbf{\Pi_2}$ are $\sin(\gamma) |0\ket + i \cos(\gamma)  |1\ket$ and $\cos(\gamma) |0\ket - i \sin(\gamma)  |1\ket$, and the eigenstates of  $\mathbf{\Pi_3}$ are $\sin(\gamma) |0\ket - i \cos(\gamma)  |1\ket$ and $\cos(\gamma) |0\ket + i \sin(\gamma)  |1\ket$.
%%= i \sqrt{(1-\lambda_1^2) (1-\lambda_2^2}/2$ turns out to be beneficial. 
%A geometrical reasoning underlying this choice and the whole approach is given elsewhere. 
The candidate density matrix  $\chi$, which can be $\rho$ or $\sigma$, is first updated using the % following 
 formula
 \vspace{-.18cm}
 \begin{eqnarray}
\mathbf{ \chi}_{i}= 
\frac{\mathbf{\pi}_i\mathbf{\chi} \mathbf{\pi}_i^\dagger}{tr(\mathbf{\pi}_i \mathbf{\chi} \mathbf{\pi}_i^\dagger)}
\nonumber
\end{eqnarray}
% \vspace{-.18cm}
 with $\mathbf{\Pi}_i= \mathbf{\pi}_i^\dagger \mathbf{\pi}_i$. %for a three component POVM, i.e. $i\in{1,2,3}$. %% and which sum up to the identity operator, i.e., $\mathbf{I}=\sum_{i=1}^3\mathbf{\Pi_i}$.  
 The  particular decomposition of $\mathbf{\Pi}_i$ is of no concern, since in subsequent calculations  $\mathbf{\pi}_i^\dagger$ and $\mathbf{\pi}_i$ always appear as a product. 
 The new cost function for a standard optimal measurement following the  POVM is 
  \vspace{-.18cm}
\begin{eqnarray}
\sum_{i=1}^N {\rm Tr}\Big( \mathbf{\Pi }_i\big(\xi \rho + (1-\xi) \sigma\big)\Big) \Big(\frac{1}{2} -\frac{1}{2}  \sqrt{1-4\xi_i (1-\xi_i) \cos^2{\delta_i}}\Big)\nonumber
\end{eqnarray} 
with
 \vspace{-.18cm}
\begin{eqnarray}
\xi_i (1-\xi_i) \cos^2{\delta_i}:= \xi(1-\xi)
 \frac{{\rm Tr}(  \mathbf{\Pi}_i \rho ){\rm Tr}( \mathbf{ \Pi}_i \sigma)}
{({\rm Tr}(\xi  \mathbf{\Pi}_i \sigma+ (1-\xi) \mathbf{\Pi}_i \rho))^2}  \frac{{\rm Tr}( \mathbf{\Pi}_i \rho \mathbf{ \Pi}_i \sigma)}{{\rm Tr}(  \mathbf{\Pi}_i \rho ){\rm Tr}( \mathbf{ \Pi}_i \sigma)}.\nonumber
\end{eqnarray} 
% The new cost function relying on the standard optimal routine subsequent to the three part intermediary measurement is
%\begin{eqnarray}
%\sum_{i=1}^{3}Tr\Big( \Pi_i (\xi \rho + (1-\xi) \sigma)\Big) \Big(\frac{1}{2} -\frac{1}{2}  \sqrt{1-4\xi_i (1-\xi_i) \cos^2{\Delta_i}}\Big).\nonumber
%\end{eqnarray} 
%In this section after some preliminaries the first steps in the calculation are carried out without yet fixing the various degrees of freedom necessary to obtain a counterexample to the Helstrom bound.
%The aim is to distinguish two special selected candidate states $\sigma$ and $\rho$, associated with priors $\xi$ and $1-\xi$ respectively, and a transition probability of $\cos^2(\delta)$ with  a cost  below the Helstrom bound  \cite{helstrom}
%\begin{eqnarray}
%\frac{1}{2} -\frac{1}{2}  \sqrt{1-4\xi (1-\xi) c^2{\delta}},\nonumber
%\end{eqnarray} 
%which relies on a single measurement round with a two component POVM.
%The two states lie on a great circle connecting the Poles $|0\ket \bra 0|$ and $|1\ket \bra 1|$.
%The transition probability is 
 As a comparison, the Helstrom cost is
  \vspace{-.18cm}
 \begin{eqnarray}
 \frac{1}{2} -\frac{1}{2}  \sqrt{1-4\xi (1-\xi) \cos^2{\delta}}.\nonumber
 \end{eqnarray} 
As a reminder,  all positive semi-definite matrices $\mathbf{A} \& \mathbf{B}$ satisfy $ {\rm Tr}(\mathbf{A B} ) \leq {\rm Tr}(\mathbf{A}) {\rm Tr}(\mathbf{B})$.  In addition, $({\rm Tr}(\mathbf{A})-{\rm Tr}(\mathbf{B}))^2\geq 0$, and therefore 
\begin{eqnarray}
 \big(   {\rm Tr}(\mathbf{\Pi}_i \rho) +
 {\rm Tr}(\mathbf{\Pi}_i \sigma)\big)^2\geq 4{\rm Tr}(\mathbf{\Pi}_i \rho){\rm Tr}(\mathbf{\Pi}_i \sigma).
\nonumber
\end{eqnarray}
For the subsequent calculation   $\xi$ is fixed at $1/2$. 
Under this condition,  there exists $\rho, \sigma$, and a set 
$\{ \mathbf{\Pi}_i\}$ such that for all $i$ the difference 
\begin{eqnarray}
 {\rm Tr}( \rho  \sigma) {\rm Tr}(\mathbf{\Pi}_i \rho){\rm Tr}(\mathbf{\Pi}_i \sigma) - {\rm Tr}(\mathbf{\Pi}_i \rho\mathbf{\Pi}_i \sigma)  \nonumber
\end{eqnarray}
is strictly positive\footnote{ Heuristic argument: 
Take each $\mathbf{\Pi}_i$ to be constructed out of $M$, with $M\ll N$,  one dimensional orthogonal subspaces %$\epsilon
 $ |e_{ki} \ket \bra e_{ki} |$  with $k\in\{1,...,M\}$ with the   inner-product of $\bra e_{ji}|e_{ki} \ket =0 $ for $k$ not equal to $j$. The quadratic in $M$ array of real parts of the products $\bra e_{ji}| \phi \ket \bra \phi |  e_{ki}\ket \bra e_{ki}|  \psi \ket \bra \psi |  e_{ji}\ket$ can be chosen to have  positive and negative outcomes. If, for example, the individual signs of the products  are randomly distributed and the absolute value of the products are all of similar size, then the sum would grow sub-linearly with the number of elements in the array.   As a consequence, for a specially chosen set $\{ \mathbf{\Pi}_i\}$, ${\rm Tr}(\mathbf{\Pi}_i \rho\mathbf{\Pi}_i \sigma)$ for each $i$ can be  smaller than   ${\rm Tr}(\mathbf{\Pi}_i \rho){\rm Tr}(\mathbf{\Pi}_i \sigma)
 $ times a selected  `constant'. 
 }. % being of a fixed value smaller than one
%. }. 
This is all one needs to breach the Helstrom bound.

%If one can show that for for all $i$ from $1,...,N$ the difference  
%\begin{eqnarray}
%{\rm Tr}(\mathbf{\Pi}_i \rho \mathbf{\Pi}_i \sigma) 
%-
%{\rm Tr}(\rho  \sigma) \big(\xi {\rm Tr}(\mathbf{\Pi}_i \rho) 
%+(1-\xi){\rm Tr}(\mathbf{\Pi}_i \sigma)\big)^2
%\nonumber
%\end{eqnarray}
%is satisfied ...
%Simplify, by setting $\xi=1/2$ %and $Tr(\mathbf{\Pi}_i \rho) =Tr( \mathbf{\Pi}_i \sigma) $
%the above difference becomes
%\begin{eqnarray}
%&&{\rm Tr}(\mathbf{\Pi}_i \rho \mathbf{\Pi}_i \sigma) 
%-
 %\frac{1}{2}  {\rm Tr}(\rho  \sigma)\big(  {\rm Tr}(\mathbf{\Pi}_i \rho) +
 %{\rm Tr}(\mathbf{\Pi}_i \sigma)\big)^2
%\nonumber\\
%&&
%\leq {\rm Tr}(\mathbf{\Pi}_i \rho \mathbf{\Pi}_i \sigma) 
%-
%{\rm Tr}(\rho  \sigma)  {\rm Tr}(\mathbf{\Pi}_i \rho) 
% {\rm Tr}(\mathbf{\Pi}_i \sigma), \nonumber
%\end{eqnarray}
%which can be negative, if $Tr(\rho  \sigma)$ has a fixed value, and $Tr(\mathbf{\Pi}_i \rho)$ have $M$ small components for each $i$ such that ....

%\noindent 
\noindent 
%The above calculation has shown that the Helstrom bound for  binary quantum state discrimination  can be breached.  The method  described  applies to states with a range of overlaps and priors. % circumvented  the paper has shown
  %, and should find application in the area of quantum algorithm. 

Should linearity be  taken for granted in quantum mechanics, is a question intimately linked with the method employed. % naturally arises, 
 Does it hold sway unconditionally? Might not linearity be only another good approximation?
If it is acknowledged that linearity is a %n another
 `god that failed', then a wholesale reassessment of quantum mechanics, and with it state discrimination, follows.

A speculative comment about a possible interpretation of quantum mechanics rounds of the appendix. %It   %is influenced 
It is related  to the long-running dispute between  views associated with   Heraclitus and Parmenides.
The claim is that change and therefore experience %in quantum mechanics 
is due to the modification of the boundary between objects\footnote{A curious example of a shift of boundary can be found in human development. An infant until a few months old, it is claimed, does not perceive the feet to be under its conscious control. The boundary eventually shifts and envelops the limbs.
%In the same vein, mental diseases could sometimes be associated with an inappropriate choice of boundary.
}.  It could  be termed the `Red Queen interpretation' of quantum mechanics, since the Red Queen  in Lewis Carroll's `Through the Looking-Glass'  commands Alice to perform ``all the running you can do, to keep in the same place. If you want to get somewhere else, you must run at least twice as fast". Unitary evolution keeps relative distances of states constant, and only the ceaseless modification of the boundary between system and environment leads to experience. %This occurs for example, when a photon is absorbed by the eye.
%\vspace{.130288cm}
%\twocolumngrid
%\vspace{-.28cm}
%\vspace{-.098cm}
%\subsection{Comments}
%\vspace{-.398cm}

\end{document}